\documentclass[nohyperref]{article}

\usepackage{microtype}
\usepackage{graphicx}
\usepackage{booktabs} 
\usepackage{url}
\usepackage{caption}
\usepackage{subcaption}
\usepackage{multirow}

\usepackage{hyperref}

\usepackage[accepted]{icml2022}

\usepackage{amsmath}
\usepackage{amssymb}
\usepackage{mathtools}
\usepackage{amsthm}

\usepackage[capitalize,noabbrev]{cleveref}

\theoremstyle{plain}

\theoremstyle{definition}

\theoremstyle{remark}

\usepackage[textsize=tiny]{todonotes}

\icmltitlerunning{Submission and Formatting Instructions for ICML 2022}

\begin{document}

\twocolumn[
\icmltitle{Don't Pay Attention to the Noise: Learning Self-supervised \\ Representations of Light Curves with a Denoising Time Series Transformer}



\icmlsetsymbol{equal}{*}

\begin{icmlauthorlist}
\icmlauthor{Mario Morvan}{ucl}
\icmlauthor{Nikolaos Nikolaou}{ucl}
\icmlauthor{Kai Hou Yip}{ucl}
\icmlauthor{Ingo P. Waldmann}{ucl}

\end{icmlauthorlist}

\icmlaffiliation{ucl}{Department of Physics and Astronomy,
University College London,
Gower Street, London, WC1E 6BT, UK}

\icmlcorrespondingauthor{Mario Morvan}{mario.morvan.18@ucl.ac.uk}

\icmlkeywords{Self-supervised Learning, Sequential Models, Time Series, Light Curves, Noise Correction, Anomaly Detection}

\vskip 0.3in
]



\printAffiliationsAndNotice{}  

\begin{abstract}
Astrophysical light curves are particularly challenging data objects due to the intensity and variety of noise contaminating them.
Yet, despite the astronomical volumes of light curves available, the majority of algorithms used to process them are still operating on a  per-sample basis.
To remedy this, we propose a simple Transformer model --called Denoising Time Series Transformer (DTST)-- and show that it excels at removing the noise and outliers in datasets of time series when trained with a masked objective, even when no clean targets are available.
Moreover, the use of self-attention enables rich and illustrative queries into the learned representations.
We present experiments on real stellar light curves from the Transiting Exoplanet Space Satellite (TESS), showing advantages of our approach compared to traditional denoising techniques\footnote{Our code is publicly available on \href{https://github.com/mariomorvan/Denoising-Time-Series-Transformer}{GitHub}.}.
\end{abstract}

\section{Introduction}

Time series of observed flux --so called `light curves'-- are one of the most common data products of space observation.
Their analysis enables the precise study of distant objects and phenomena within and beyond the solar system and the Milky Way including stars \citep[e.g.][]{christensen-dalsgaard_asteroseismology_2007}, planets \citep[e.g.][]{charbonneau_detection_2000, di_stefano_possible_2021} asteroids \citep{warner_asteroid_2009} or black holes \citep[e.g.][]{beskin_detection_2002}. 
However, light curves are often affected by instrumental, photon and background noise. In addition, the target itself often shows an undesirable variability 
of similar frequencies to the underlying  scientific signal, making an optimal noise filter difficult to achieve. All these factors render the analysis of light curves challenging, often requiring technical expertise to build specialised pre-processing pipelines before physical modelling and interpretation. 

Although the use of deep learning has started to emerge to successfully address some problems related to light curves \citep[e.g.][]{sarro_automatic_2006,wang_causal_2016,hlozek_results_2020,shallue_identifying_2018,pearson_searching_2018, morvan_detrending_2020, nikolaou_lessons_2020}, these often address only the later stages of data analysis and are limited to building supervised learning models. These models are indeed generally trained on scarcely labelled or simulated data and thus suffer from biases or small training sizes when applied to new or full datasets.
On the other hand, there already exist large datasets consisting of thousands to billions of light curves \citep[e.g.][]{bakos_wide-field_2004, pollacco_wasp_2006, auvergne_corot_2009, butters_first_2010, borucki_kepler_2010} with many more being generated by existing and future space telescopes.
We believe that tailored deep learning models will be able to leverage these large datasets to improve the efficacy and efficiency of light curve processing in a self-supervised, semi-supervised or unsupervised manner.

The self-attention mechanism \citep{parikh_decomposable_2016} and the Transformer architecture \citep{vaswani_attention_2017} have initiated a revolution in the field of natural language processing \citep[e.g.][]{devlin_bert_2019, NEURIPS2020_1457c0d6} and later computer vision \citep{khan_transformers_2021}. Transformers exhibit good generalisation, and offer easier training and better scalability compared to Long Short-Term Memory Networks \citep{hochreiter_long_1997}.
Work is under way to adapt the Transformer architecture for time series tasks such as forecasting \citep[e.g.][]{li_enhancing_2020, zhou_informer_2021, woo_etsformer_2022}.
In their study \citet{zerveas_transformer-based_2021} successfully pre-trained a Time Series Transformer via a masked objective before fine-tuning it for classification and regression. 
Even though the use of masked objectives is common in the aforementioned works, here our main objective is to denoise the time series. The masked objective allows us to solve the problem by means of a proxy imputation task without requiring any fine-tuning.

%
Our main contributions consist in: 
(i) introducing a simple self-supervised framework to perform time series denoising without access to clean targets; 
(ii) demonstrating how a Transformer encoder with minimal modification can perform light curve denoising effectively, leveraging the number and diversity of available inputs,
(iii) producing flexible\footnote{The flexibility of the model lies in its capability to handle inputs with missing values, variable sizes and generating processes characterised by different variances.} and interpretable predictions by visualising attention scores associated with imputation and denoising of sequences. 

We present experiments on real light curves from the Transiting Exoplanet Survey Satellite  \citep[TESS,][]{ricker_transiting_2015}. 
This is the first time a deep learning model is proposed to try to address both imputation and denoising on a dataset of light curves.


\section{Methodology}
\subsection{Problem Formulation}
Given a univariate time series $x=\{x_1,..,x_t,..,x_T\} \in \mathbb{R}^T$ we seek to predict its \textit{trend}\footnote{The `trend' here can contain low frequency variability, aperiodic or periodic patterns.} $y \in \mathbb{R}^T$ which has been corrupted by a noise process $\epsilon$ such as $x_t = y_t + \epsilon_t$ for each time step $t$.
No assumption is made about the corruption process except its independence from the trend. In particular, $\epsilon$ can be heteroscedastic and non-Gaussian.

Let us consider a generic model $f$ solely fed with corrupted time series, i.e. trained without clean targets in a \textit{Noise2Self} setting \citep{batson_noise2self_2019}. 
After masking a fraction of each input $x$ with randomly generated masks $m$, $f$ produces predictions $\hat{y}=f(x)$ of the same length as the input but is trained with a regression loss computed solely on the masked values: $\mathcal{L}(f(x), x, m)$.
This masked objective guarantees the independence of the predictions with respect to the local values and their associated noise. 
If missing values are present in the dataset, they are treated in the same way as randomly masked values, making the method robust to missing values. The only difference is that predictions for truly missing values are not included in the calculation of the training loss.




\subsection{Denoising Time Series Transformer}
\begin{figure}[ht]

\begin{center}
\includegraphics[width=\columnwidth]{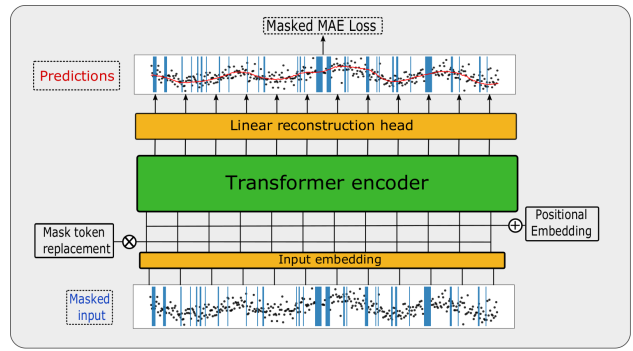}
\end{center}
\caption{Schematic overview of the DTST model learning trend representations of inputs with a masked objective. Masks represented in shaded blue areas include both missing and randomly masked time steps during training. At test time, only truly missing values are masked. Yellow modules represent the time-distributed linear embedding and the prediction head respectively.}
\label{fig:model-overview}
\end{figure}

An overview of the DTST is shown on Figure \ref{fig:model-overview}. 
For each input time series an input mask is generated combining missing values and artificially masked values (at training only). 
Masked and standardised inputs are linearly projected into input embeddings $ z \in	\mathbb{R}^{T\times D} $ of the model's dimension $D$. 
Input embeddings corresponding to masked positions are replaced by a learnable vector of dimension $D$, inspired by the mask token used in \citet{devlin_bert_2019}. This is a robust way of informing the model of the masked input positions.  Additionally, we have found a learnable vector to perform better than replacing masked values by zero as is often done for time series imputation models \citep[e.g.][]{cao_brits_2018, zerveas_transformer-based_2021, yi_why_2020}. 
However using this scheme on its own affects the quality of the predictions outside the masks and for this reason we replace $10\%$ of masked values in input with uniformly sampled values between $-2$ and $2$ and do not replace their projected input embedding by the mask embedding. This setting was the most effective we have tried amongst several ones listed in Appendix \ref{appendix:mask}.

Positional embeddings are then added to the input embeddings to provide positional information. Since they produced better results than trainable positional embeddings in our experiments we used the same fixed positional embeddings as in \citet{vaswani_attention_2017}.
We use a light version of the original Transformer encoder with the hyperparameters fixed to the values shown in Appendix\,\ref{appendix:hyperparams}. Each encoder's output is finally projected back into the input dimension using a distributed $D\times1$ linear layer.

\section{Experiments} 
\subsection{Dataset}


We present experiments on a dataset of light curves from the TESS satellite, acquired during the first visit of its first sector in 2018. 
TESS light curves are challenging because of their length ($20,076$ time steps for short cadence data spread over $30$ days), their noise level, residual instrumental systematics and missing blocks.

We select $2$ minutes cadence light curves at the Presearch Data Conditioning stage, i.e. after removal of the main instrument systematics, cosmic rays and background noise with the standard TESS pipeline \citep{jenkins_tess_2016}.
After rejection of 50 light curves with negative flux, the dataset contains 15839 light curves. We selected $20\%$ of all light curves uniformly at random for testing and the remaining $80\%$ for training and validation.


\subsection{Training and evaluation}
Because of their length we randomly crop each light curve to select $400$ consecutive time steps. 
A random mask is then generated before subtracting the mean and dividing by the standard deviation of the non-masked values for each input segment. This procedure can be seen as a data augmentation step, as the combination of cropping and masking operations will produce different inputs at each epoch.  

For training the DTST we use the noise-scaled masked mean absolute error (NMMAE) defined as:
$\text{NMMAE}(\hat{y}, x, m) = \frac{1}{M*n(x)} \sum_{t=1}^T{m_{t}|x_{t}-\hat{y}_{t}|}$,
\noindent where $m$ is a binary mask equal to $1$ for masked time steps and $0$ otherwise, $\hat{y}$ is the model's output prediction, $M=\sum_{t=1}^T{m_{t}}$ is the total number of masked steps in $x$ and $n(x)$ is an estimate of the local noise by computing the average moving standard deviation with a window of size 10 and a step of 5. 
Compared to the mean-squared error, the mean absolute error (MAE) is more robust to outliers while rescaling using $n(x)$ helps to account for different variabilities in the training data.
Predictions for the full light curves are then obtained by stitching together the predictions for segments of 400 time steps. In practice, evaluation segments are designed so as to allow overlaps of 50 steps and remove the outer 25 steps for each prediction. 

As evaluation metrics we use the MAE and the inter-quartile range (IQR) of the detrended light curve as a measure of the residual noise, both expressed as percentages of the stellar flux. For both measures, lower values are desirable.

We compare the DTST to the median filter and Tukey's biweight algorithms with implementations from \citet{hippke_wotan_2019} as baselines. These have shown optimal or near-optimal performance in removing the noise prior to detecting exoplanets in Kepler and TESS data. Both methods require to set the window length --in time units for Tukey's algorithm and in number of cadences for the median filter. For comparison we select two window lengths: a long window of $6$ hours ($\sim 300$ time steps) which is adapted for exoplanet transit detection and a short window of $\sim 2$ hours which provides comparable denoising scores to the DTST but overfits some of the high frequency variability.




\subsection{Results}

After experimenting with various architectures and masking scenarios (see appendix \ref{appendix:mask}) on the training set, we evaluated the DTST and the baselines on the test set. Results are presented in Table \ref{table:results}. 
On average, the DTST provides the smallest residual noise and auto-correlation out of the several baselines evaluated here. 
The difficulty for traditional techniques here lies in reconciling the diversity of the stellar processes composing the dataset, and it is therefore understandable that a single cadence-based or window-based filter with a fixed window size will either fail to denoise targets with high variability or overfit the noise on those with low variability.

\begin{table}[ht]
\caption{Denoising performance on 3168 test light curves from Sector 1. Averaged errors are given in percentage of the stellar flux. Window sizes considered by the three algorithms to make predictions are shown on the second line.}
\small
\label{table:results}
\label{results}
\begin{center}
\begin{tabular}{lccccc}
    & \multicolumn{2}{c}{\textbf{MEDIAN FILTER}} &   \multicolumn{2}{c}{\textbf{BIWEIGHT}}   &  \textbf{DTST}\\
 & $65$ steps & $181$ steps & $2$ h & $6$ h & $400$ steps \\
\hline \\

IQR                 &   $0.393\%$       &  $0.465\%$   &   $0.398\%$   &    $0.469\%$            &   {\boldmath$0.385\%$}   \\
MAE                 &  $0.244\%$  &     $0.286\%$ & $0.245\%$           & $0.286\%$           &   {\boldmath\textbf{$0.235\%$}}
\end{tabular}
\end{center}
\end{table}

\begin{figure*}[!ht]
     \centering
     \begin{subfigure}[b]{0.41\linewidth}
         \centering
         \includegraphics[width=\textwidth]{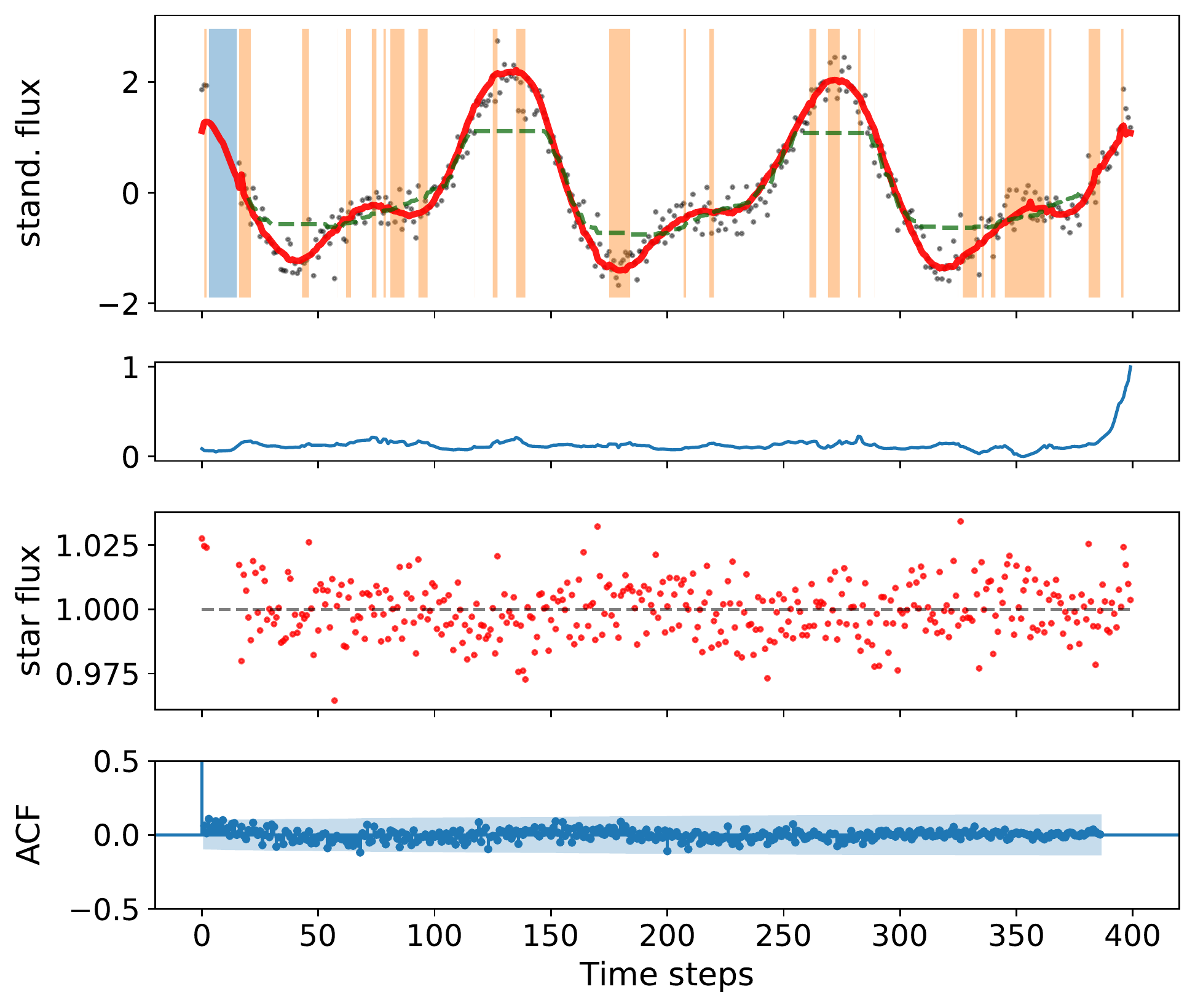}
         \caption{TESS Target 260504446}
         \label{fig:pred-a}
     \end{subfigure}
     \hfill
     \begin{subfigure}[b]{0.41\linewidth}
         \centering
        \includegraphics[width=\textwidth]{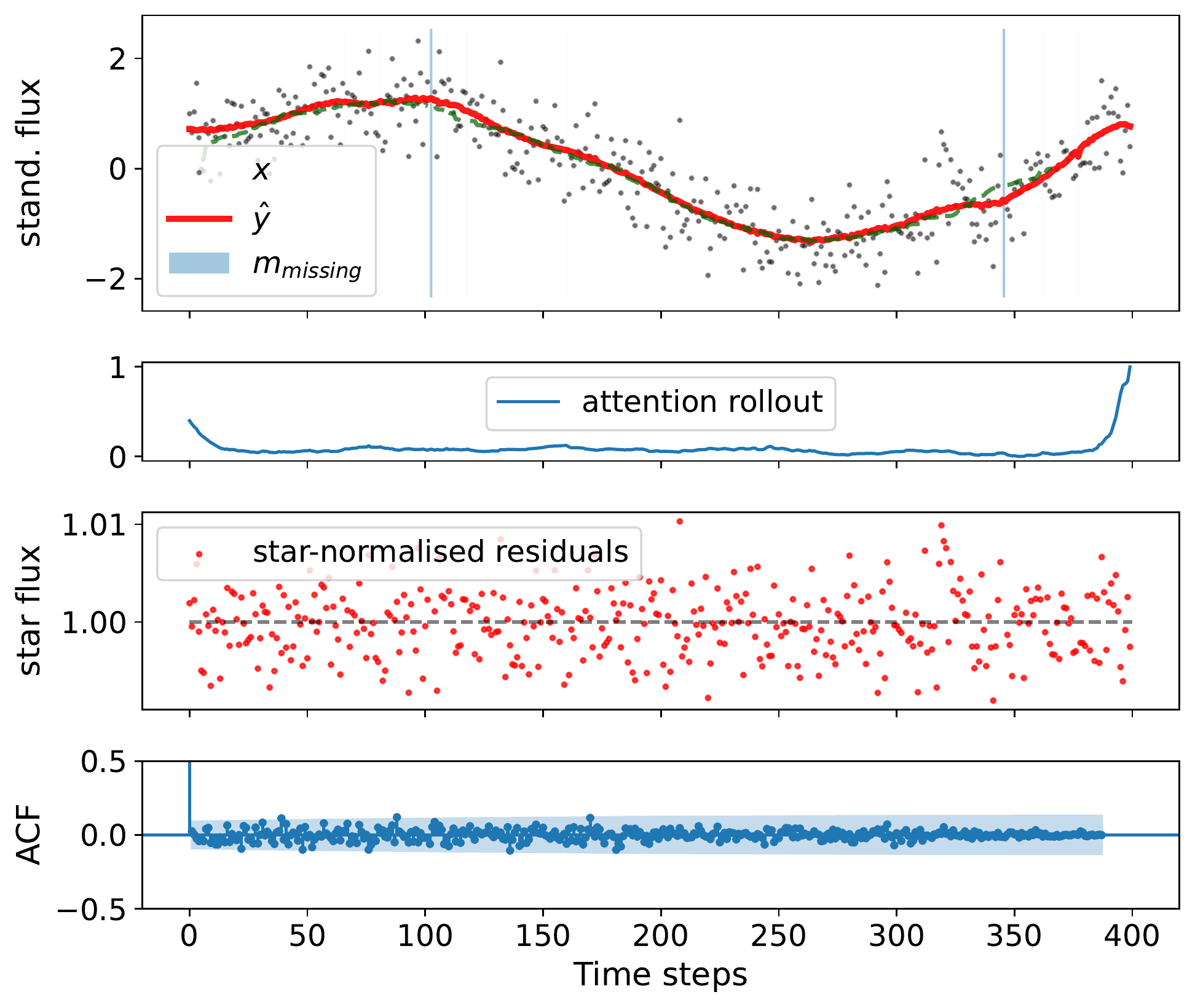}
        \caption{TESS Target 269829656}
         \label{fig:pred-b}
     \end{subfigure}
     \hfill
     \begin{subfigure}[b]{0.41\linewidth}
         \centering
        \includegraphics[width=\textwidth]{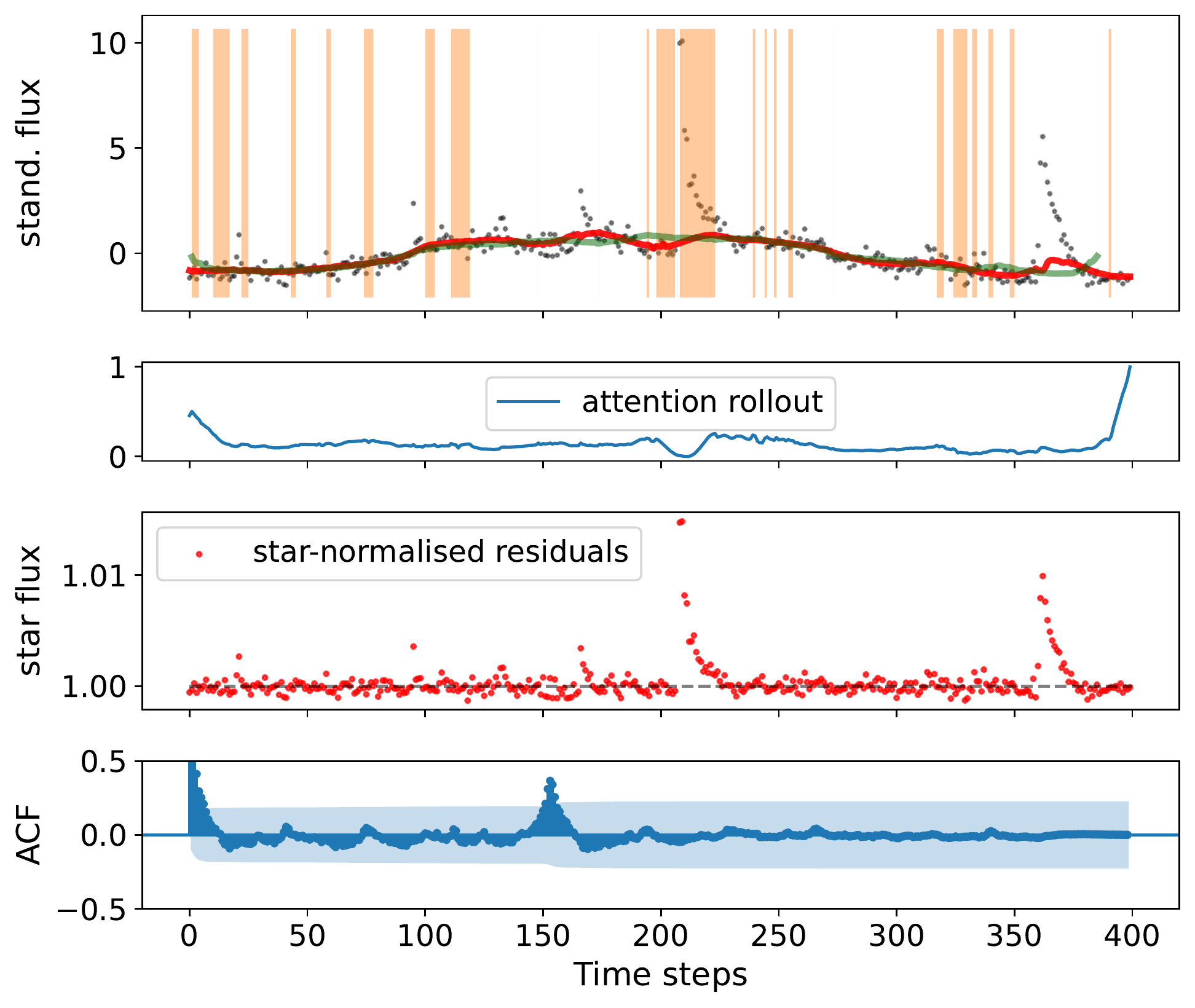}
        \caption{TESS Target 140045538}
         \label{fig:pred-c}
     \end{subfigure}
     \hfill
     \begin{subfigure}[b]{0.41\linewidth}
         \centering
         \includegraphics[width=\textwidth]{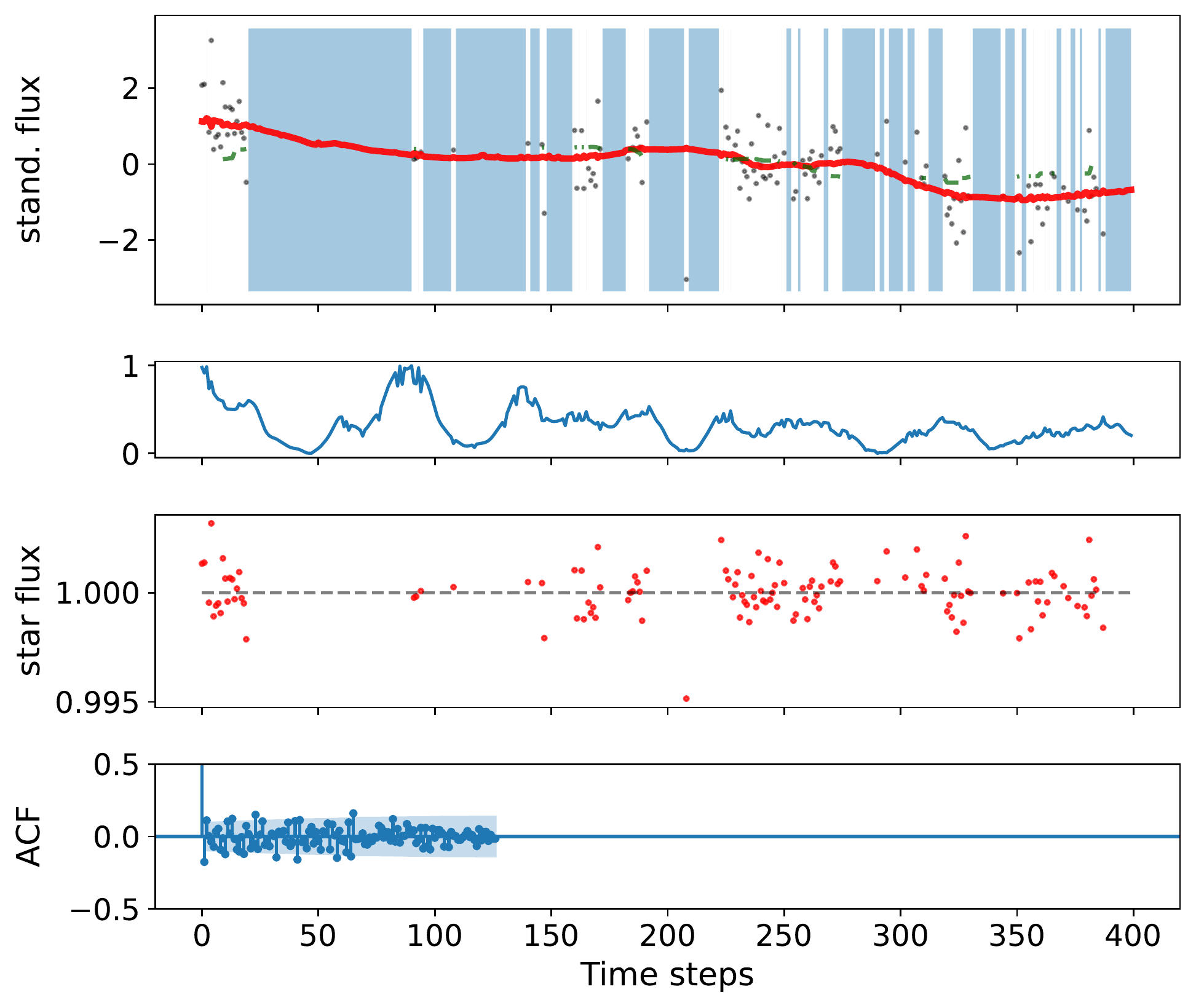}
        \caption{TESS Target 92257800}
         \label{fig:pred-d}
     \end{subfigure}
\caption{Diagnosis of predictions for four different test stars. On the left are two examples with random artificial masks (in orange) mimicking the training process. On the right are two uncorrupted inputs, where the blue shaded masks indicate truly missing data in input. Each sub-figure contains from top to bottom: (i) inputs as black dots,  the DTST's predictions as red line, and median filter with window of $65$ cadences as green dashed line, (ii) rolling attention time series scaled between $0$ and $1$, (iii) the star-normalised residual errors and (iv) the auto-correlation function with missing data ignored.}
\label{fig:predictions}
\end{figure*}

We show examples of predictions in Figure \ref{fig:predictions} for different test samples showing a range of variability patterns. 
We corrupted the two inputs on the left (Figures \ref{fig:pred-a} and \ref{fig:pred-b}) with random masks similar to those used during training.
The predicted time series shown in red on each upper sub-plot plot shows very good agreement with the expected trend for both masked and unmasked input time steps. In dashed green line is shown a the result of a median filter on each light curve with a window of 65 cadences (equivalent to ~2 hours). While it provides good results for slowly varying stellar processes (Figure \ref{fig:pred-b}), this setting fails to account for faster processes (Figures \ref{fig:pred-a} and \ref{fig:pred-c}) or inputs with many missing values (Figure \ref{fig:pred-d}).

On each third sub-plot we show the residual light curve in units of stellar flux. 
The associated autocorrelation function (ACF) of the model-fit residual is plotted on the last subplot of each figure.
The ACF is a useful tool to analyse the significance of residual time correlations as in Figure \ref{fig:pred-c}. Target 140045538 indeed shows bursts of flaring activity which are not predicted by the DTST and therefore leave a significant signature in the ACF.
As short transients or planetary transits may lie at the border between noise, outliers and signals, further fine-tuning of the model may be needed for either predicting or ignoring them consistently.



\subsection{1D Attention Maps}
We use Rolling Attention \citep{abnar_quantifying_2020} to combine the attention scores of all layers and heads. We direct the reader to Appendix \ref{appendix:attention} for more details and examples.
This enables us to visualise which parts of the inputs received more attention for producing the outputs, both during training and validation.
Thus we are using the generated attention maps both for orienting the model's development and interpreting its predictions.

Our first observation is that both input tails often receive high rolling attention scores. This is understandable as these lack context on either their left or right and therefore prove more challenging to predict. 
We also observed that large masked regions receive generally less attention than non masked regions. This is in fact a useful check during the model's development to verify if the model manages to distinguish between the mask representation and the real values. Furthermore, values surrounding the identified gaps often show greater attention than average, probably as they are particularly relevant for the prediction of masked values. Finally it is often interesting to look at the attention patterns for time steps corresponding to unexpected flux values. Those are sometimes ignored such as the rightmost flaring event on Figure \ref{fig:pred-c} or conversely receive more attention than average when they can inform predictions. 




\section{Conclusion}

In this work we presented a conceptually simple framework to denoise time series via a proxy imputation task. We performed experiments and showed how such an approach based on a Transformer encoder architecture is effective at removing the noise in light curves from the TESS satellite. 
Compared to traditional techniques, this model can offer flexibility and increased performance when pre-processing large datasets of light curves.
Further works will extend these experiments to other real and simulated datasets while assessing the generalisation power and possible gain from using a pre-trained model. 
Finally we would like to use this approach as a basis for downstream tasks such as event detection, imputation and upsampling.





\bibliography{main.bib}
\bibliographystyle{icml2022}

\newpage
\appendix
\onecolumn
\section{1D Attention Maps}
\label{appendix:attention}
\begin{figure}[ht]
\begin{center}
\includegraphics[width=0.95\linewidth]{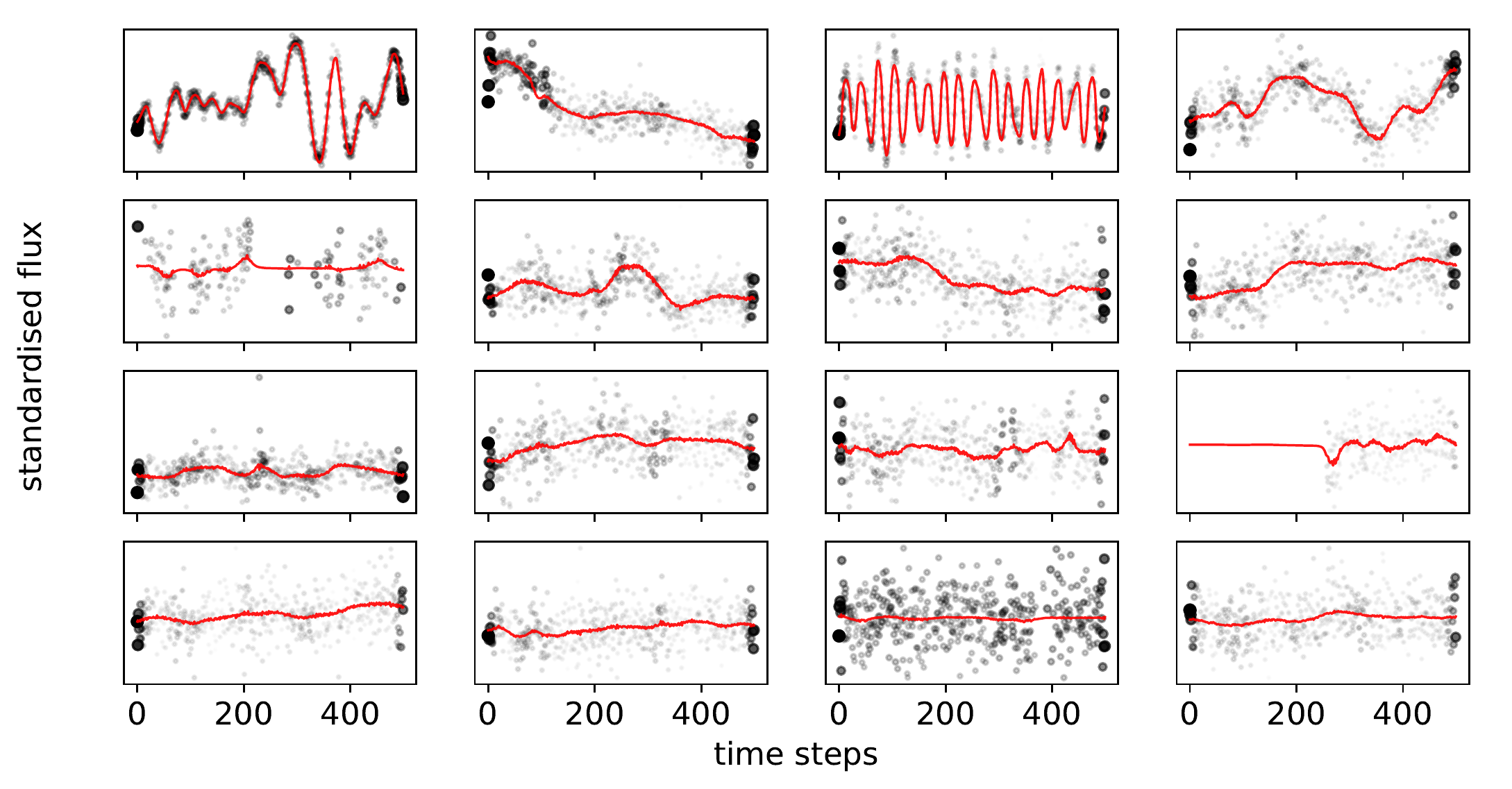}
\end{center}
\caption{1D attention maps computed with Attention Rollout and overlayed with predictions (red lines) for 16 random light curves from TESS dataset. The size and opacity of individual point inputs is directly proportional to the Attention Rollout values.}
\label{attention_preds}
\end{figure}

Attention from the model's output to the input time series is computed using Attention Rollout \citep{abnar_quantifying_2020}. This procedure consists in recursively multiplying matrices of attention weights through the transformer layers, thus accounting for mixing of attention in the network. Figure \ref{attention_preds} shows more examples of predictions overlayed with their corresponding input with Rollout Attention used to highlight time steps with greater attention. 

\section{Masking Strategy}
\label{appendix:mask}
\paragraph{Masking Patterns} We explored various generating mechanisms for the distributions of masked values in each input. Given a fixed ratio of values to mask, we  tested using a Bernoulli distribution for time-independent masking and a geometric distribution over masked blocks lengths. The geometric distribution was used to impose longer masks and thus a more challenging imputation task to the model. Mean block lengths of 5, 10 and 20 were tested and final results were presented for a window of $10$ as it offered the best compromise between the length of signals to impute and denoising performance.Intuitively, the length of masked blocks will control the degree of temporal locality of the noise processes to remove, and using wider masked regions will indeed force the model to make use of longer-term dependencies to make accurate predictions.

\paragraph{Masking Ratios}
We set the masking ratio to $30\%$ after experimenting with $10\%$, $20\%$, $30\%$, $40\%$ and $50\%$. Heuristically, increasing the masking ratio speeds up training but also affects performance as fewer inputs are available for prediction.
When values are missing in the inputs, the masking ratio is considered with respect to the number of non-missing time steps. This maintains the ratio of data used for training constant while avoiding degenerate cases where the random mask would be empty or would cover the entirety of the non-missing input.

\paragraph{Replacement Strategy}
We considered various replacement strategies for  masked input values: 
(i) replacing by zero, 
(ii) by a uniformly random value centred on zero, 
(iii) by a special learnable vector \citep[inspired by ][]{devlin_bert_2019} in the model's space, and 
(iv) keeping the original values.   
Case (iv) was quickly discarded as it led to overfitting the noise. 
While option (iii) offered the best imputation performance, we observed that it performed poorly on its own for denoising the full inputs, and that this issue was mitigated by using case (ii) for a random fraction of input time steps, even as small as $10\%$. This can be understood as an extra corruption operation on the input, thus forcing the model to provide coherent predictions even outside the regions whose embeddings are more explicitly masked with a dedicated vector. Whilst we have compared these several cases, it would be interesting to investigate further the influence of the replacement strategy (e.g. different distributions) and ratios on the denoising performance.

\section{Hyperparameters}
\label{appendix:hyperparams}

Fixed hyperparameters for all presented experiments are shown in Table \ref{hyperparams}.

For training we used Adam \citep{kingma_adam_2015} optimiser with learning rate $0.001$ and $\beta=(0.9, 0.999)$.

\begin{table}[ht]
\caption{Hyperparameters}
\label{hyperparams}
\begin{center}
\begin{tabular}{lc}
\textbf{PARAMETER}           & \textbf{VALUE}     \\
\hline \\
Learning rate       & $0.001$   \\
Batch size          & $64$     \\
Dim. model          & $64$     \\
Dim. feedforward    & $128$     \\
Num layers          & $3$       \\
Num. heads          & $8$       \\
Masking ratio       & $30\%$    \\
Average masking length & $10$     \\
\end{tabular}
\end{center}
\end{table}

\section{Computational Efficiency}
Even though training the DTST on thousands of time series can take up to several hours on a single V100 GPU, its inference cost remains very low with around $10~\mu s$ for a full TESS light curve unfolded in windows of length $400$ passed as a batch of size $\sim60$. This is to be compared with $\sim50~\mu s$ and $\sim173~\mu s$ per TESS light curve for the efficient Wotan implementations of biweight and median filter respectively.

The $O(T^2)$ complexity in space and time of vanilla attention could be mitigated by using sparse attention (see e.g. \citet{zhou_informer_2021} in $O(T\log T)$, \citet{wang_linformer_2020} in $O(T)$). 
Additional experimental studies would need to be performed to evaluate their respective impact on performance and on the trade-off between long sequences and full attention.



\end{document}